\documentclass[letterpaper, 10 pt, conference]{ieeeconf}  

\IEEEoverridecommandlockouts                              

\usepackage{amsmath,amssymb,enumerate,soul}
\usepackage{epstopdf}
\usepackage{times} 
\usepackage[dvips]{graphicx}
\usepackage{psfrag}
\usepackage{multirow} 
\usepackage{cite} 
\usepackage{float}
\usepackage[x11names]{xcolor}
\usepackage{colortbl}
\usepackage{mathtools} 
\usepackage{booktabs} 

\usepackage{algorithm}

\usepackage{algpseudocode}
\usepackage{pifont}

\usepackage{amsthm}
\theoremstyle{plain}
\newtheorem{assmp}{Assumption}

\newtheorem{thm}{Theorem}

\newtheorem{rmrk}{Remark}
\newtheorem{defn}{Definition}
\newtheorem{lma}{Lemma}

\theoremstyle{definition}

\newcommand{\R}[2]{\ensuremath{\mathbb{R}^{#1}_{#2}}}
\newcommand{\N}{\ensuremath{\mathbb{N}}}
\newcommand{\Np}{\ensuremath{\mathbb{N}_{\geqslant 1}}}
\newcommand{\NT}{\ensuremath{\mathbb{N}_{\geqslant T}}}

\newcommand{\MAX}{\ensuremath{\textnormal{MAX}}}

\newcommand{\T}{\mathcal{T}}
\newcommand{\barV}{\overline{\mathcal{V}}}
\newcommand{\barE}{\overline{\mathcal{E}}}
\newcommand{\barN}{\overline{\mathcal{N}}}

\usepackage{color}
\newcommand{\jmh}[1]{{#1}}
\providecommand{\quotes[1]}{``#1"}

\newcommand{\MA}[1]{{#1}}

\title{\Huge MAX-consensus in open multi-agent systems with gossip interactions}

\author{\Large Mahmoud Abdelrahim, Julien M. Hendrickx and W.P.M.H. Heemels
\thanks{Mahmoud Abdelrahim is with the Department of Mechanical Engineering, Assiut University, Egypt.  Julien Hendrickx is with the ICTEAM institute, Universit\'{e} catholique de Louvain, Louvain-la-Neuve, Belgium. Maurice Heemels is with the Control Systems Technology Group, Department of Mechanical Engineering, Eindhoven University of Technology, the Netherlands. Email addresses: {\tt\small\fontfamily{txtt}\selectfont m.abdelrahim@aun.edu.eg}, {\tt\small\fontfamily{txtt}\selectfont julien.hendrickx@uclouvain.be}, {\tt\small \fontfamily{txtt}\selectfont m.heemels@tue.nl}.}
\thanks{Mahmoud Abdelrahim and Maurice Heemels were supported by the Innovational Research Incentives Scheme under the VICI grant ``Wireless control systems: A new frontier in automation'' (No. 11382) awarded by NWO (The Netherlands Organisation for Scientific Research) and STW (Dutch Technology Foundation). Julien Hendrickx's work is partially supported by the Belgian Network DYSCO (Dynamical Systems, Control, and Optimization), funded by the Interuniversity Attraction Poles Program, initiated by the Belgian Science Policy Office, and by the  Concerted  Research  Action  (ARC)  programme  supported  by
the Federation Wallonia-Brussels.}}

\begin{document}
\maketitle

\begin{abstract}
We study the problem of distributed maximum computation in an open multi-agent system, where agents can leave and arrive during the execution of the algorithm. The main challenge comes from the possibility that the agent holding the largest value leaves the system, which changes the value to be computed. The algorithms must as a result be endowed with mechanisms allowing to forget outdated information. The focus is on systems in which interactions are pairwise gossips between randomly selected agents. We consider situations where leaving agents can send a last message, and situations where they cannot. For both cases, we provide algorithms able to eventually compute the maximum of the values held by agents.
\end{abstract}

\section{Introduction} \label{sec: Introd}

Multi-agent systems involve interacting elements with computing capabilities, also called agents or nodes, who communicate with each other to achieve a collective control task that is \jmh{more} difficult or \jmh{sometimes even} impossible to be performed by an individual agent. This configuration of multi-agent systems has a great benefit to model and solve many problems in different fields of applications including sensor networks \cite{Olfati-ACC08-Distributed, Shi-IET10-Resource, Demigha-CST12-On}, computer networks \cite{Cerf-TC74-protocol, Muthuk-TCS98-First} and social science \cite{Hegs-JASSS02-opinion, Blondel-SIAM10-continuous, Liu-CDC12-Dynamic}. One of the common problems that has been studied in these applications is the consensus of multi-agent systems on aggregate functions such as, e.g., MIN, MAX, SUM and AVERAGE. For instance, in a group of distributed sensors, it can be required to compute the average temperature of a specific region or to elect the sensor with maximum power resource to preserve the communication over a costly link or to reduce energy for a wireless sensor network, see, e.g., \cite{Boyed-TAC06-Randomized, Iutz-IEEETSP12-max-xonsensus, Giannini-TCS16-Asynch}.

Most existing results of the literature rely on the assumption that the system composition is static, i.e., the set of agents present in the system does not change after the  initial time, see, e.g., \cite{Olfati-IEEE07-Consensus, Iutz-IEEETSP12-max-xonsensus, Hendrickx-TAC11-Distributed, Ren-ACC05-Survey} and the references therein. However, this requirement can be difficult to satisfy in some implementation scenarios where new agents can join and/or existing agents can leave the network at any time instant. This phenomenon is known in the literature as ``\textit{network churn}" \cite{Stutz-ACM06-Understanding, Kuhn-DistComp10-Towards}, ``\textit{dynamic network}" \cite{Jelas-ACM05-Gossip, Kuhn-ACM10-Distributed, Dutta-ACM13-On} or ``\textit{open multi-agent systems}" \cite{Hendrickx-AAC16-Open, Huynh-AAMAS06-An, Pinyol-AIR13-Computational}. In this case, the consensus problem on aggregate estimates becomes more challenging to handle compared to the case of static networks. For instance, consider the paradigmatic problem of MAX-consensus with distributed communications and assume that the agent with the largest state value has left the network after all the agents have converged to its state value. In this case, all the existing nodes in the network will then hold outdated information.
This scenario cannot occur in static networks, which highlights one of the inherent challenges of open multi-agent systems.

In this paper, we investigate the problem of MAX-consensus in open multi-agent systems with distributed communications. The agents are assumed to be anonymous, do not have global identifiers, and  all run the same algorithm.
We further assume that interactions only occur via pairwise \quotes{gossip} exchanges between randomly selected agents in the sense that, at any (discrete) time instant, (only) two agents are selected randomly to exchange their information,  update their MAX estimates and possibly other variables. To cope with the dynamic nature of the network, two different solutions are proposed depending on whether or not it is possible for the leaving agents to announce their departures.

\jmh{In the case in which announcements are made, our algorithm relies on a variable that describes how \quotes{up-to-date} agents are with respect to recent departures, and priority is given to information coming from the most \quotes{up-to-date} agents. In the case where agents disappear without sending a last message, our algorithm maintains an estimate of the age of the information, and estimates corresponding to information deemed too old are discarded.}

We will show that our two approaches ensure that outdated information can be forgotten, and that the consensus on the MAX value can be achieved (with high probability) if the system composition stops evolving.

The problem of MAX-consensus in multi-agent systems has been studied in, e.g., \cite{Nejad-ISICAT09-MAx, Iutz-IEEETSP12-max-xonsensus, Zhang-IEEESJ16-MAX, Giannini-TCS16-Asynch}. Among existing techniques, the work of \cite{Iutz-IEEETSP12-max-xonsensus} has considered MAX-consensus with random gossip interactions between agents, \jmh{on a \textit{static} network.}
\MA{Compared to existing works of the literature, our result is adapted to the problem of MAX-consensus in multi-agent systems when the network is \emph{open}, which has not been considered in the previously mentioned works. Our proposed approach encompasses the result of pairwise gossip interaction in static networks in \cite{Iutz-IEEETSP12-max-xonsensus} as a particular case.}

The remainder of the paper is organized as follows. Notations are given in Section \ref{sec: notation}. The problem is formulated in Section \ref{sec: problem-formulation}. In Section \ref{sec: counters}, we treat the case where leaving agents send a last message, and in Section \ref{sec: timeout}, we treat the case where they do not.
Numerical simulations are given in Section \ref{sec: example}. Conclusions \jmh{and discussions} are provided in Section \ref{sec: conclusion}.

\section{Notation} \label{sec: notation}
Let $\R{}{} := (-\infty,\infty)$, $\R{}{\geqslant 0} := [0,\infty)$, $\N := \{ 0, 1, 2, \ldots \}$, $\Np := \{ 1, 2, \ldots \}$ and $\NT := \{ T, T+1, \ldots \}$ for $T\in\N$. We denote by $\boldsymbol 0_n$ and $\boldsymbol 1_n$ the vectors in $\R{n}{}$ whose all elements are 0 or 1, respectively. We write $A^{T}$ to denote the transpose of $A$, and $(x,y)\in\R{n_x+n_y}{}$ to represent the vector $[x^{T}, y^{T}]^{T}$ for $x \in \R{n_x}{}$ and $y \in \R{n_y}{}$. The symbol $\mathbb{I}_{n}$ stands for the identity matrix of dimension $n$. For a random variable $R$, the symbol $\mathbb{E}(R)$ denotes the expectation of $R$.

\section{Problem statement} \label{sec: problem-formulation}

Consider a connected time-varying graph $\mathcal{G}(t)=(\mathcal{V}(t), \mathcal{E}(t))$, where $\mathcal{V}(t)$ and $\mathcal{E}(t)$ denote, respectively, the set of existing agents and the set of edges in the graph at time $t\in\N$. The graph $\mathcal{G}(t)$ is dynamic in the sense that new agents can join and/or existing agents can leave at any time $t$. Hence, the cardinality of $\mathcal{V}(t)$, denoted by $\mathcal{N}(t)$, is not necessarily constant for all $t\in\N$. The agents communicate with each other in a \textit{pairwise} randomized gossip fashion \cite{Boyed-TAC06-Randomized}. \jmh{In other words, at any time instant $t\in\N$, there are three possibilities: (i) an agent joins the system and $\mathcal{N}(t+1)= \mathcal{N}(t)+1$, (ii) an agent leaves the system and  $\mathcal{N}(t+1)= \mathcal{N}(t)-1$, or (iii) two randomly selected agents $i,j\in\mathcal{V}(t), i\neq j$ communicate with each other (Note that these discrete-time instants may be interpreted as the sampling of an asynchronous process at those times where an event occurs). Joining agents are assumed to know that they join the system. Leaving agents may or may not be able to send one last message (to one other agent) before leaving, which are two cases of interest, which will be discussed in Section \ref{sec: counters} and \ref{sec: timeout}}, respectively.

Every agent $i$ has \jmh{two special states: $x_i\in \R{}{}$ is its intrinsic value, which is constant and determined arbitrarily when joining the system, and $y_i(t)$ is its estimated answer at time $t\in\N$ for the MAX value. Our goal is to estimate the maximum intrinsic value of all the agents present in the system,} \MA{so we would ideally want, when no more agents are joining or leaving the network after time $T\in\N$, that there is a time $T^*\in\N\geqslant T$ such that $y_i(t) = \MAX(t) := \max_{j\in \mathcal{V}(t)} x_j$, for all $i\in \mathcal{V}(t)$ and for $t\in\N\geqslant T^*$.} Agents may then have other states that they use to reach this goal.

If the network would be static, i.e., $\mathcal{G}(t)$ is time-invariant, the estimation of the maximum could be achieved in finite time by starting from $y_i(0)=x_i(0)$ for every agent, and setting $y_i(t+1)=y_j(t+1)=\max(y_i(t),y_j(t))$ whenever agents $i$ and $j$ interact at time $t\in\N$, see, e.g., \cite{Iutz-IEEETSP12-max-xonsensus}. The main challenge in a dynamic or open network lies with the need for the algorithm to take new agents into account and to eventually discard information related to agents no longer present in the system to ensure that $\max_jx_j$ is eventually recovered once the system composition stops evolving. \MA{Classical algorithms such as that in \cite{Iutz-IEEETSP12-max-xonsensus} do not guarantee this: outdated values from agents no longer in the system may never be discarded.}

\jmh{Note that an alternative and maybe more natural goal would be to have the $y_i(t), i\in\mathcal{V}(t)$ track $\MAX(t) = \max_{j\in \mathcal{V}(t)} x_j$ sufficiently accurately. This more ambitious goal is left for future studies, see Section \ref{sec: conclusion} for further discussions on this issue.}

Finally, we chose to make the following assumption for the sake of simplicity of exposition.

\MA{
\begin{assmp} \label{assmp:graph-complete}
The graph $\mathcal{G}(t)=(\mathcal{V}(t), \mathcal{E}(t))$ is complete for all $t\in\N$. \hspace*{\fill} $\Box$
\end{assmp}}
This means that every pair of distinct agents in the network can communicate directly with each other. The algorithms we develop would actually also work on general dynamic graphs under suitable connectivity assumptions, but the analysis would be more complex.

\section{Departures are announced} \label{sec: counters}

\subsection{Algorithm description}
If leaving agents announce their departure (to one other agent), then we can benefit from this knowledge to correct the outdated information. For that purpose, we introduce an auxiliary variable $\kappa_i(t)\in\N$ at each agent, meant to represent the \quotes{level of information} available to $i$ about the  departures up to time $t\in \N$. It will in general \emph{not be equal to the actual  number of departures, nor converge to it}. The algorithm is designed to ensure that those with the largest value $\kappa_i$ have valid estimates, i.e. their $y_i(t)$ correspond to the $x_j$ of agents present in the system. For this purpose, information coming from agents with higher $\kappa_i$ will be given priority over information coming from agents with lower values, and it will be made sure that agents with a lower value of $\kappa_i$ will never have influenced those with a higher value.

The algorithm is summarized as follows. Initially, every existing agent at $t=0$ sets $y_i(0) = x_i$ and $\kappa_i=0, i\in \mathcal{V}(0)$, as shown in Algorithm \ref{Intialization algorithm}.
\begin{algorithm}[h]
\caption{Intialization algorithm}
\label{Intialization algorithm}
At time $t=0$, every existing node $i\in\mathcal{V}(0)$ initializes its state as
\begin{algorithmic}[1]
\State $y_i(0) = x_i$
\State $\kappa_i(0) = 0$
\end{algorithmic}
\end{algorithm}

When a new agent $n$ joins the group at time $t\in\N$, it initializes its counter $\kappa_n(t)$ and its estimate $y_n(t)$ according to Algorithm \ref{Joining algorithm}.
\begin{algorithm}[h]
\caption{Joining algorithm}
\label{Joining algorithm}
Assume at any time $t\in\Np$, a new agent $n$ wants to join, i.e., $\mathcal{V}(t+1) = \mathcal{V}(t)\cup\{n\}$. Agent $n$ initializes its state as
\begin{algorithmic}[1]
\State $y_n(t) = x_n$
\State $\kappa_n(t) = 0$
\end{algorithmic}
\end{algorithm}

If an agent $\ell$ leaves the system, it sends a last message containing its counter value $\kappa_\ell$ to a randomly selected agent $m$. The reaction of $m$ is governed
by Algorithm \ref{Departure algorithm}, which can be interpreted as follows: If the counter $\kappa_\ell(t)$ of the leaving agent $\ell$ is less than $\kappa_m(t)$, then agent $m$ ignores this departure since the information of $\ell$ is deemed less up-to-date than its own and has not influenced it. 
On the other hand, if $\kappa_\ell(t) \geqslant\kappa_m(t)$, then $\ell$ may have influenced $m$ and possibly agents $i$ with values $\kappa_i$ higher than $\kappa_m$, but no larger than $\kappa_{\ell}$. To ensure that none of the agents with the highest values $\kappa$ hold the now outdated value $x_l$,
$m$ will reset its $y_m$ to $x_m$, which is by definition a valid value, and set its $\kappa_m$ to $\kappa_{\ell}+1$, a value above that of all those who could have been influenced by $\ell$.

\begin{algorithm}[h]
\caption{Departure algorithm}
\label{Departure algorithm}
 Assume at time $t\in\N$, agent $\ell$ leaves, i.e., $\mathcal{V}(t+1) = \mathcal{V}(t)\setminus\{\ell\}$.\\
 Agent $\ell$ picks a random agent $m$ to inform, and agent $m$ updates its state as follows
\begin{algorithmic}[1]
\If{$\kappa_\ell(t) < \kappa_m(t)$}
\State $y_m(t+1) = y_m(t)$
\State $\kappa_m(t+1) = \kappa_m(t)$
\ElsIf{$\kappa_\ell(t) \geqslant \kappa_m(t)$}
\State $y_m(t+1) = x_m$
\State $\kappa_m(t+1) = \kappa_\ell(t) + 1$
\EndIf
\end{algorithmic}
\end{algorithm}

The gossip communication between agents is performed via Algorithm \ref{Modified random gossip} (values not explicitly updated remain constant between $t$ and $t+1$). When $\kappa_i(t) = \kappa_j(t)$, this implies that agents $i$ and $j$ either have not been informed about any departure from the group, i.e., $\kappa_i(t) = \kappa_j(t) = 0$, or have equal information level about the departure of one or more agents. In either case, agents $i$ and $j$ can exchange their information to update their estimate for the MAX value. When $\kappa_i(t) > \kappa_j(t)$, agent $i$'s information about past departures is deemed more up to date. Agent $j$ is then not allowed to transfer its estimate $y_j$ to avoid infecting $i$ with possibly outdated information (unless its estimate is actually its own value, which is by definition valid). Therefore, agent $j$ restarts to $\max(y_i(t), x_j)$ and increments its counter to $\kappa_j(t+1) = \kappa_i(t)$ in order to alert other future agents who have not been informed yet to restart. The case when $\kappa_j(t) > \kappa_i(t)$ is completely symmetric.

\begin{algorithm}[h]
\caption{Gossip algorithm}
\label{Modified random gossip}
At each time step $t\in\N$, two agents $i,j\in\mathcal{V}(t)$ are picked randomly (with possibly $i=j$)
\begin{algorithmic}[1]
\If{$\kappa_i(t) = \kappa_j(t)$}
\State $y_i(t+1) = y_j(t+1) = \max(y_i(t), y_j(t))$
\ElsIf{$\kappa_i(t) > \kappa_j(t)$}
\State $y_i(t+1) = y_j(t+1) = \max(y_i(t), x_j)$
\State $\kappa_j(t+1) = \kappa_i(t)$
\Else
\State $y_i(t+1) = y_j(t+1) = \max(x_i, y_j(t))$
\State $\kappa_i(t+1) = \kappa_j(t)$
\EndIf
\end{algorithmic}
\end{algorithm}

\subsection{Eventual Correctness} \label{sec:correct_with_messages}

\jmh{We now show that the algorithm described in the previous subsection is correct in the sense that, with high probability (and even almost surely), it eventually settles on the correct value if arrivals and departures stop.}

Remember that $\mathcal{V}(t):=\{1,\ldots,\mathcal{N}(t)\}$ denotes the group of agents \jmh{present at time $t$, and let $X:=\{x_1,\ldots,x_{\mathcal{N}(t)}\}$ be the set } of intrinsic values of nodes in $\mathcal{V}(t)$. Assume that after some time $T\in\N$ no agent leaves and no new agent joins the system, so that $\mathcal{V}(t) = \mathcal{V}(T)=\barV$, $\mathcal{E}(t) = \mathcal{E}(T)=\barE$ and $\mathcal{N}(t)= \mathcal{N}(T)=\barN$  for all $t\in\NT$. Then, we need to show that all the currently existing agents $\mathcal{V}(T)$ in the network will successfully reach the correct maximum value. For that purpose, we define the following property.

\MA{
\begin{defn}\label{def: eventual-correctness}
We say that an algorithm is eventually correct if for any $T\in\N$ with $\mathcal{G}(t)=(\barV, \barE)$ for all $t\in \N_{T}$, there exists a $T^* \in\NT$ such that $y_i(t) = \max_{j\in  \overline{\mathcal{V}}}x_j$ for all $i\in\barV$ and all $t\in\N_{T^*}$.
\end{defn}
}

Denote $K(t):=\underset{i\in\mathcal{V}(t)}{\max}\kappa_i(t)$, MAX$(t)=\underset{i\in\mathcal{V}(t)}{\max}x_i$ and $X_{K}(t):=\{x_i: i\in\mathcal{V}(t) \wedge \kappa_i(t)=K(t)\}$. We state the following result.
\begin{lma} \label{lma: max-kappa}
  For all $t\in\N$ and any $j\in\mathcal{V}(t)$, if $\kappa_j(t)=K(t)$ then $y_j(t)\in X_{K}(t) \subseteq X(t)$.
\end{lma}
\noindent Lemma \ref{lma: max-kappa} states that, at any time $t\in \N$, if the counter value of an agent $j\in\mathcal{V}(t)$ is equal to the maximum value $K(t)$, then its estimate $y_j(t)$ is equal to an intrinsic value $x_i\in X_{K}(t)$ of one of the agents present in the system at this time $t$ and whose value $\kappa_i$ is $K(t)$. 

\vspace{0.5cm}
\noindent \textbf{Proof.} Consider any agent $j\in \mathcal{V}(t)$ with $\kappa_j(t)=K(t)$. We have three scenarios:

\noindent (a) Agent $j$ has just joined the \jmh{system} at time $t$. Hence, $\kappa_j(t)=0$ and $y_j(t)=x_j$ according to Algorithm \ref{Joining algorithm}. Since $\kappa_j(t)= K(t)$, this implies that $K(t)=0$. Hence, $K_i(t)=0$ for all $i\in\mathcal{V}(t)$. Consequently, it holds that $y_j(t)\in X_{K}(t) = X(t)$. ~\\[-2pt]

\noindent (b) $K(t)>K(t-1)$ \jmh{(and $j$ is not a new agent)}. In this case, since at most one agent can change its counter at any time, there is exactly one agent $j$ with $\kappa_j(t) = K(t)$. This implies that an agent $\ell\in\mathcal{V}(t-1)$ with $\kappa_\ell(t-1)=K(t-1)$ has left at time $t$ and informed agent $j$ about its departure, otherwise $\kappa_j(t)\neq K(t)$ or $K(t)\ngtr K(t-1)$. Consequently, agent $j$ restarts according to lines 7-9 in Algorithm \ref{Departure algorithm} and we have that $\kappa_j(t)=K(t-1)+1=K(t)$ and $y_j(t)=x_j\in X_{K}(t)$. ~\\[-2pt]

\noindent (c) $K(t)=K(t-1)$ \jmh{(and $j$ is not a new agent)}. We have two possibilities:

\noindent c1)    
$\kappa_j(t)>\kappa_j(t-1)$, i.e., agent $j$ has increased its counter value at time $t$ such that $\kappa_j(t)=K(t)=K(t-1)$, which can happen by one of the following actions:
    \begin{itemize}
      \item [-] an agent $i\in\mathcal{V}(t-1)$ with $\kappa_i(t-1) = K(t-1)-1\geqslant \kappa_j(t-1)$ has left the group and informed agent $j$ about its departure. Consequently, in view of lines 4-6 in Algorithm \ref{Departure algorithm}, agent $j$ has incremented its counter to $\kappa_j(t)=\kappa_i(t-1)+1=K(t-1)=K(t)$, otherwise $\kappa_j(t)\neq K(t-1)$ and $y_j(t)=x_j\in X_K(t)$.
      \item [-] no departure occurred but agent $j$ has interacted with an agent $i\in\mathcal{V}(t-1)$ with $\kappa_i(t-1)=K(t-1)$. Consequently, in view of lines 4-6 in Algorithm \ref{Modified random gossip}, we obtain $\kappa_j(t)=\kappa_i(t-1)=K(t-1)=K(t)$ and $y_j(t)=\max(x_j, y_i(t-1))\in X_{K}(t)$.
    \end{itemize}~\\[-2pt]

\noindent c2) $\kappa_j(t)=\kappa_j(t-1)$, i.e., agent $j$ did not increase its counter value at time $t$. Then, since $K(t)=K(t-1)$, it holds that $\kappa_j(t-1)=K(t-1)$ and we know that $y_j(t-1)=x_i$ for some $x_i\in X(t-1)$. There are two different possibilities:
    \begin{itemize}
    \item [-] $y_j(t)\neq y_j(t-1)$, which can only happen if agent $j$ has interacted via algorithm \ref{Modified random gossip} with an agent $h$ with $\kappa_h(t-1)= K(t-1)$. Hence, in view of line 3 in algorithm \ref{Modified random gossip}, it holds that $y_j(t)=y_h(t-1)\in X_{K}(t)$.
    \item [-] $y_j(t) = y_j(t-1) = x_i$. We know that agent $i$ did not leave because otherwise it would have been true that agent $i$ has informed some neighbour $m$ about its departure and resulted in $\kappa_m(t)=\kappa_i(t-1)+1 = K(t-1)+1$, which leads to case (b) not case (c). Hence, since $i\in\mathcal{V}(t)$, it holds that $y_j(t)\in X_{K}(t)$.
  \end{itemize}
 This completes the proof of Lemma \ref{lma: max-kappa}. \hspace*{\fill} $\Box$

\MA{
\begin{thm} \label{thm: convergence}
Suppose that Assumption \ref{assmp:graph-complete} holds. Then, Algorithm \ref{Intialization algorithm}-\ref{Modified random gossip} is eventually correct.
\end{thm}
}

\vspace{0.5cm}
\noindent \textbf{Proof.}
The proof of Theorem \ref{thm: convergence} relies on Lemma \ref{lma: max-kappa} and the result developed in \cite{Iutz-IEEETSP12-max-xonsensus}. Note that an essential difference between our problem and the setup in  \cite{Iutz-IEEETSP12-max-xonsensus} is that the gossip interaction between agents (as in Algorithm \ref{Modified random gossip}) depends considerably on their counter values, which is not the case in static networks as in \cite{Iutz-IEEETSP12-max-xonsensus}. Therefore, we will invoke their result twice, once on the counter values $\kappa_i$ to show that all agents eventually have the maximal counter value $K(T)$, and once on the actual estimate $y_i(t)$ to show that they eventually reach $\MAX(T)$.

\jmh{After time $T$, only Algorithm \ref{Modified random gossip} is applied. Ignoring for the moment its effect on the $y_i(t)$, observe that it performs a classical gossip operation on the $\kappa_i(t)$, in the sense that an interaction between $i$ and $j$ results in $\kappa_i(t+1) = \kappa_j(t+1) = \max(\kappa_i(t),\kappa_j(t))$.} Theorem 4, 5 in \cite{Iutz-IEEETSP12-max-xonsensus}, applied to complete graphs following Assumption \ref{assmp:graph-complete}, allows us then to guarantee that the counters of all agents converge to the maximum counter value $K(T)$ in a finite time $T^*_1$ with the following properties
\begin{equation}\label{T1star-expect}
 \mathbb{E}(T^*_1-T) \leqslant (\barN -1)h_{\barN-1},
\end{equation}
where $h_n$ denotes the $n$th harmonic number, i.e., $h_n:=\sum_{k=1}^{n}\frac{1}{k}$. Moreover, we have, with probability $1-\epsilon$ that $T^*_1-T$ is bounded by
\begin{equation}\label{T1star-bound}
(\barN -1)h_{\barN-1}
  \left(1 \!+\! \log\left(\frac{\barN}{\epsilon}\right)\left(1 \!+\! \sqrt{1 \!+\! \frac{1}{\log\frac{\barN}{\epsilon}}}\right)\right).
\end{equation}

\jmh{After $T$, since $\kappa_i(t)=K(t)$ for all $i$, it follows from Lemma \ref{lma: max-kappa} that all $y_i(t)$ correspond to actual values $x_j, j\in \barV$. Moreover, since one can easily verify that $y_i(t)\geqslant x_i$ at all times, there holds $\max_{i\in \barV}y_i(t) = \max_{i\in \barV}x_i=  \MAX(t)=\MAX(T)$. It is therefore sufficient to show that all $y_i(t)$ eventually settle on the same value.}

For this purpose, observe that when all agents have the same $\kappa_i(t) = K(t)$, Algorithm \ref{Modified random gossip} reduces to its line 2, $y_i(t+1)=y_j(t+1) = \max(y_i(t),y_j(t))$,  which is again a classical pairwise gossip. We can then re-invoke Theorem 4, 5 in \cite{Iutz-IEEETSP12-max-xonsensus} to show the existence of a $T^*$ after which
$y_i(t) = \max_{i\in \barV} y_i(T^*) = \max_{i\in barV} x_i=\MAX$, with the same bounds on $T^*-T^*_1$ as on $T^*_1-T$. In particular,
$\mathbb{E}(T^*-T)\leqslant2(\barN -1)h_{\barN-1}$, and there is a probability
$1-\epsilon$ that $T^*-T$ is at most twice the expression in \eqref{T1star-bound}. This achieves the proof of Theorem \ref{thm: convergence}. \hspace*{\fill} $\Box$

\begin{rmrk}
Note that, since we apply the result of \cite{Iutz-IEEETSP12-max-xonsensus} \emph{twice} to prove that Algorithm \ref{Intialization algorithm}-\ref{Modified random gossip} is eventually correct, the upper bound that we obtain on the time needed to achieve this property is conservative. This comes from the fact that, in Algorithm \ref{Intialization algorithm}-\ref{Modified random gossip}, the agents update their counters and their estimates simultaneously and not sequentially. \hspace*{\fill} $\Box$
\end{rmrk}

\section{Departures are not announced} \label{sec: timeout}

\subsection{Algorithm description}

Leaving agents may not always be able to announce their departure, such as in case of unforeseen failures or disconnections.
The algorithms in Section \ref{sec: counters} can no longer be applied in such a more challenging setting. Therefore, we now propose an alternative algorithm that does not use messages from departing agents.
The idea is to have each agent maintain a variable $\T_i$ representing the \quotes{age} of its information. This age $\T_i$ is kept at 0 when the agent's estimate $y_i(t)$ of $\MAX(t)$ corresponds to (only) its own value $x_i$, as the validity of its information is then guaranteed. Otherwise $\T_i$ is increased by 1 every time agent $i$ interacts with another agent, as the information gets \quotes{older}. When an agent $i$ changes its estimate $y_i(t)$ by adopting the estimate $y_j(t)$ of an agent $j$, it also sets $\T_i(t)$ to the value $\T_j(t)$, which corresponds to the age of the new information it now holds. Finally, when $\T_i(t)$ reaches a threshold $\T^*$, the information $y_i(t)$ is considered too old to be reliable and is discarded; $y_i(t)$ is reset to $x_i$ and $\T_i(t)$ to 0. We defer the discussion on the value of $\T^*$ to Section \ref{sec: conclusion}, but already note that it should depend on (bounds of) the system size, or (possibly) change with time.

Formally, the behavior of an agent joining the system is governed by Algorithm \ref{Joining algorithm_timeout}, while the update of $\T_i(t)$ and the gossip interactions are governed by Algorithms \ref{update_timer} and \ref{Modified_random_gossip_timeout_sep_from_update} (where we use $y_i(t^+),\T_i(t^+)$ to denote intermediate values the variables $y_i,\T_i$ may take during the computation leading to their values at $t+1$). Observe that when $i$ and $j$ have the same estimate $y_i(t)=y_j(t)$ they update the age of information to the smallest among $\T_i(t)$ and $\T_j(t)$. Observe also that the algorithms guarantee that $y_i(t)\geqslant x_i$ for every $i$ at all times, since $y_i(t)$ can never decrease except when it is re-initialized at $x_i$.
Finally, there is no algorithm for the departure, since agents are not assumed to be able to take any action when other agents leave as this is not announced.

\begin{algorithm}[H]
\caption{Joining algorithm}
\label{Joining algorithm_timeout}
Assume at time $t\in\Np$, a new agent $n$ wants to join. Agent $n$ initializes its state as follows
\begin{algorithmic}[1]
\State $y_n(t) = x_n$
\State $\mathcal{T}_n(t) = 0$
\end{algorithmic}
\end{algorithm}

\begin{algorithm}[H]
\caption{UpdateTimer}
\label{update_timer}
When agent $i$ calls this procedure\footnote{T}:
\begin{algorithmic}[1]
\If{$y_i(t) = x_i$} \Comment{guaranteed validity of estimate}
\State{$\T_i(t^+) = 0$}
\State{$y_i(t^+) = x_i$}
\Else\Comment{estimate gets one period older}
\State $\T_i(t^+)=\T_i(t)+1$
\State{$y_i(t^+) = y_i(t)$}
\EndIf
\If{$\T_i(t) = \T^*$} \Comment{Reset if threshold reached}
\State{$y_i(t^+)  = x_i$}
\State{$\T_i(t^+)=0$}
\EndIf
\end{algorithmic}
\end{algorithm}

\begin{algorithm}[H]
\caption{Gossip algorithm}
\label{Modified_random_gossip_timeout_sep_from_update}
At each time step $t$, two agents $i,j$ are picked randomly
\begin{algorithmic}[1]
\State UpdateTimer(i), UpdateTimer(j)
\If{$y_i(t^+) > y_j(t^+)$}
\State $y_j(t+1) = y_i(t^+)$
\State $\T_j(t+1) = \T_i(t^+)$
\ElsIf{$y_j(t^+)<y_i(t^+)$}
\State $y_i(t+1)=y_j(t^+)$
\State $\T_i(t+1)=\T_j(t^+)$
\ElsIf{$y_j(t^+)=y_i(t^+)}$
\State $\T_i(t+1),\T_j(t+1) := \min(\T_i(t^+),\T_j(t^+))$
\EndIf
\end{algorithmic}
\end{algorithm}

\subsection{Eventual Correctness}
We now discuss the eventual correctness of the algorithm described above. For space reasons, only sketches of proofs will be presented. We use the same conventions as in Section \ref{sec:correct_with_messages}. We first prove that outdated values are eventually discarded if agents stop leaving or arriving.

\begin{lma}\label{lma:discard_time_out}
If no arrival or departure takes place after time $T\in\N$, then almost surely there exists a time $T'\in\N$ after which every estimate $y_i$ corresponds to the value of an agent present in the system, i.e., for $t\in\N\geqslant T'$, for all $i$ there exists a $j\in \mathcal{V}(t)=\overline{\mathcal{V}}$ such that $y_i(t) = x_j$. As a consequence, $y_i(t) \leqslant\MAX(T) = \max_{j\in \overline {\mathcal{V}}} x_j$ for every $i\in \mathcal{V}(t)=\overline{\mathcal{V}}$ and $t\in\N\geqslant T'$.
\end{lma}

\noindent \textbf{Proof.} 
Observe first that agents can only set their $y_i$ to their own $x_i$ or to the value $y_j$ of some other agent. Hence, since the set of values $x_i$ remains unchanged after $T$, values $y_i(t)$ for times $t\geqslant T$ that are not equal to some $x_j, j\in \overline{\mathcal{V}}$, must be equal to some $y_j(T)$, i.e., must have been held as estimated at time $T$. We show that these outdated values are eventually discarded.

Let $z\in \R{}{}$ be such an outdated value, that is, $y_i(T) = z$ for some $i\in \barV$ but  $z= x_j$ for no $j\in \barV$. Let then $D(t)=\{i\in \barV: y_i(t)=z\}$ be the set of agents holding $z$ as estimate at time $t$, and $\tau(t) = \min\{\T_i(t):i\in D(t)\}$ be the minimal age of information at $t$ for those holding this outdated value as estimate. As long as $D(t)$ is non-empty, there must hold $\tau(t) \leqslant\T^*$ due to the reset in Algorithm \ref{update_timer}. We will show that $\tau(t)$ must keep increasing if $D(t)$ remains non-empty, leading to a contradiction.

Every time an agent $i\in D(t)$ for which $\T_i(t) = \tau(t)$ interacts with some other agent, It follows from Algorithm \ref{Modified_random_gossip_timeout_sep_from_update} and the timer update in Algorithm \ref{update_timer} that it must increase its counter $\T_i$ by 1, unless it changes its value $y_i$ and no longer belongs to $D(t+1)$. In both cases the set of agents in $D(t)$ with this $\T_i$ taking this value has decreased by 1.
Besides, since $z$ is equal to no $x_i$, the only way an agent $i$ can join $D(t+1)$ if it was not in $D(t)$ is by interacting with an agent $j\in D(t)$, and the rules of the algorithm imply then that $\T_i(t+1)= \T_j(t) +1 \geqslant \tau(t) +1$.
Hence $\tau(t)= \min_{i\in D(t)} T_i(t)$ never decreases, and when it is not increasing, the number of agents in $D(t)$ for which $\T_i(t) = \tau(t)$ either remains constant, or decreases as soon as one of them is involved in an interaction (once it reaches 0, $\tau(t)$ automatically increases). Since all agents are almost surely repeatedly involved in interactions, this means $\tau(t)$ will almost surely eventually increase as long as $D(t)$ is nonempty, in contradiction with the fact that it cannot exceed $\T^*$. $D(t)$ must thus almost surely eventually be empty, which means that any outdated value is thus almost surely eventually discarded, so that after some time $T'$ every estimate $y_i(t)$ corresponds to a $x_j$ for $j\in \barV$. \hspace*{\fill} $\Box$

Let us now prove that the agents' estimates $y_i$ eventually take the correct value $\MAX$ with a high probability.

\begin{thm}\label{thm:correctness_time_out}
For all $\epsilon>0$, there exists a (sufficiently large) $\T^*\in\N$ such that, if no arrival or departure takes place after time $T\in\N$, then there exists a time $T''\in\N\geqslant T$ after which $y_i(t) =\MAX(T) = \max_{j\in \overline {\mathcal{V}}}x_j$ holds for every $i\in \overline {\mathcal{V}}$ with a probability at least $1-\epsilon$.
\end{thm}

\noindent \textbf{Proof.}
Let $m$ be an agent holding the maximal value after time $T$: $x_m = \MAX = \max_{i\in \mathcal{V}}x_i$. It follows from Lemma \ref{lma:discard_time_out} that  $y_m(t)\leqslant\MAX$ holds after some $T'$, which implies $y_m(t) = \MAX= x_m$, since one can verify that $y_i(t) \geqslant x_i$ holds for all agents at all times. The timer update Algorithm \ref{update_timer} implies then that $\T_m(t)=0$ at all times after $T'$.

Let us now fix some arbitrary time $t_0\geqslant T'$ and let $C(t)\subseteq \overline{\mathcal{V}}$ be the set of agents $i$ such that (i) $y_i(t) = \MAX$, and (ii) $\T_i \leqslant t - t_0$. The set $C(t_0)$ contains at least agent $m$. Moreover, for $t\in [t_0,t_0+\T^*-1]$, there holds $C(t) \subseteq C(t+1)$.
Indeed, observe first that no agent of $C(t)$ \quotes{resets} because the $\T_i$ of agents in $C(t)$ are by definition smaller than $\T^*$. Moreover, agents in $C(t)$ do not change their value $y_i$ either because it follows from Lemma \ref{lma:discard_time_out} that no agent $j$ has a value $y_j\geqslant y_i= \MAX$, so condition (i) still holds.
Besides, the timer $\T_i$ increase by at most 1 at each iteration so condition (ii) also holds.
Observe now that whenever an agent $i\in C(t)$ interacts with an agent $j\not\in C(t)$ at a time $t\in [t_0,t_0+\T^*-1]$, agent $j$ will set $y_j(t+1)$ to $y_i(t) = \MAX $ and join $C(t+1)$. A reasoning similar to that the analysis of classical pairwise gossip algorithm in \cite{Iutz-IEEETSP12-max-xonsensus} shows then that, for every $\epsilon$, there exists of a $\tau$ given by \eqref{T1star-bound} such that with probability at least $1-\epsilon$, all agents will be in $C(t)$ after $t_0+\tau$ and at least until $t_0+\T$ (provided $\T \geqslant \tau$).
There would thus hold $y_i=\MAX$ for all $i$.
Since this holds true for any arbitrary $t_0\geqslant T'$, it follows that for every $i$ and $t\geqslant T'+\tau$,  $y_i(t)= \MAX$ holds with probability at least $1-\epsilon$. \hspace*{\fill} $\Box$

The proofs of eventual correctness show that the value of the threshold $\T^*$ is subject to a trade-off:
We see from the proof of Lemma \ref{lma:discard_time_out} that the time needed to discard outdated values increases when $\T^*$ is increased. On the other hand, a sufficiently large threshold is needed in Theorem \ref{thm:correctness_time_out}. In its proof, we see that larger thresholds allow larger $\tau$, which imply smaller probabilities $\epsilon$ of some agent not having the correct value.

Besides, we see in Theorem \ref{thm:correctness_time_out} that $\T^*$ must be sufficiently larger than the expression \eqref{T1star-bound}, \emph{which depends on $\barN$}, the eventual size of the system. This implies that agent must know at least a bound on this size, unlike in the algorithm developed in Section \ref{sec: counters} when leaving agents could send a last message. One theoretical solution to avoid this problem would be to let $\T^*$ slowly grow with time, so that it would eventually always be sufficiently large if the system composition stops changing (This growth should be sufficiently slow for the argument of Lemma \ref{lma:discard_time_out} still to be valid).
However, the system would also become slower and slower in discarding outdated information.

\section{Simulations} \label{sec: example}
We demonstrate the application of our algorithms on a group of 25 agents:  Initially, the intrinsic states $x_i$ of all agents were assigned to random integer values between 0 and 1000. The largest two values of $x_i$ are found to be $x_{9} = 936$ and $x_{13} = 815$. The estimates $y_i(0)$ for all agents are initialized to $x_i$ and all the counters $\kappa_i(0)$ and ages $\T_i(0)$ are initialized to 0. Agent $9$ with the highest value, $x_9=936$ leaves at $t=200$. Pairwise interactions between two randomly selected agents take place at every other time.

\begin{figure}
\centering
\begin{tabular}{c}
\includegraphics[scale=.4]{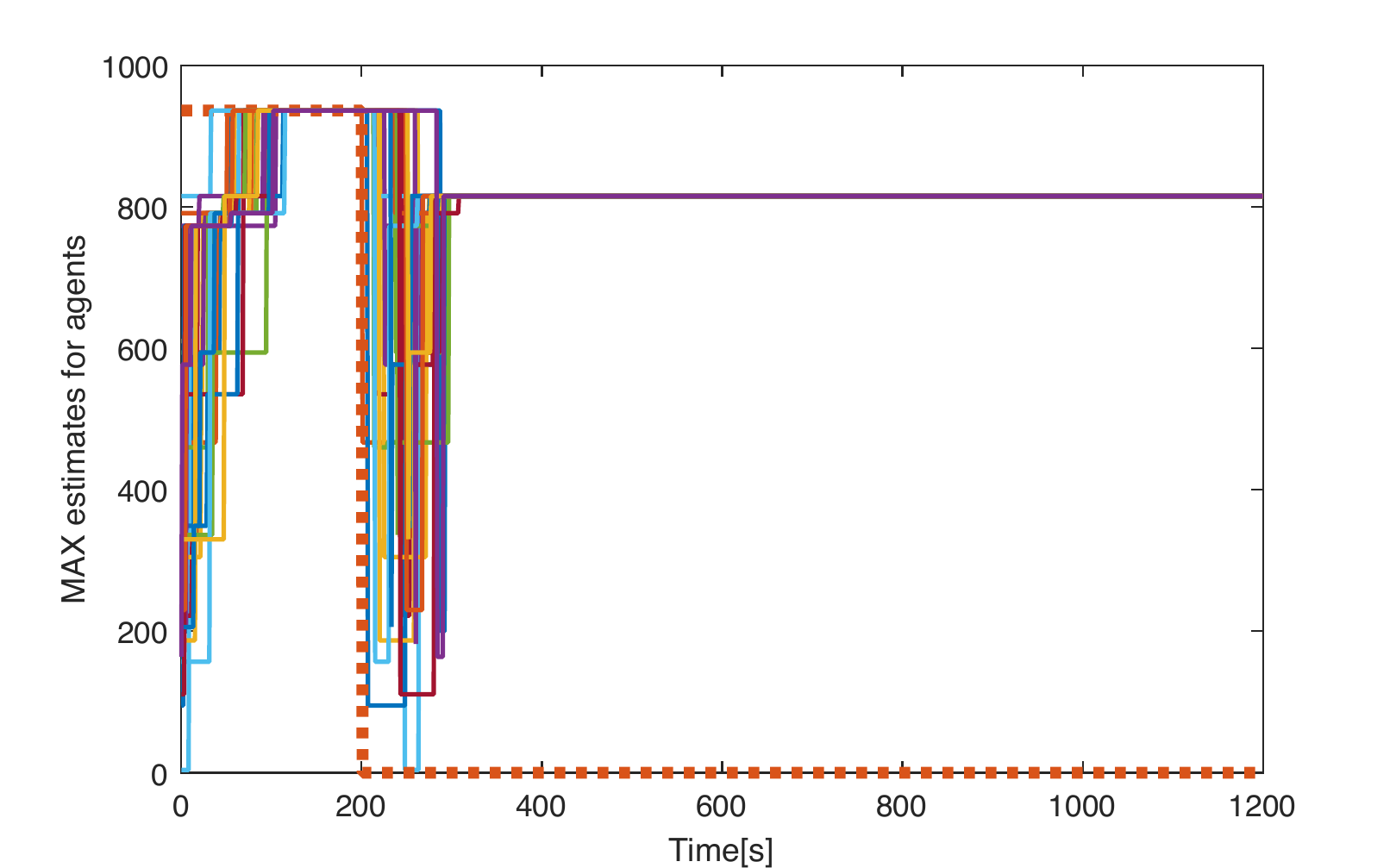}\\
(a)\\
\includegraphics[scale=.4]{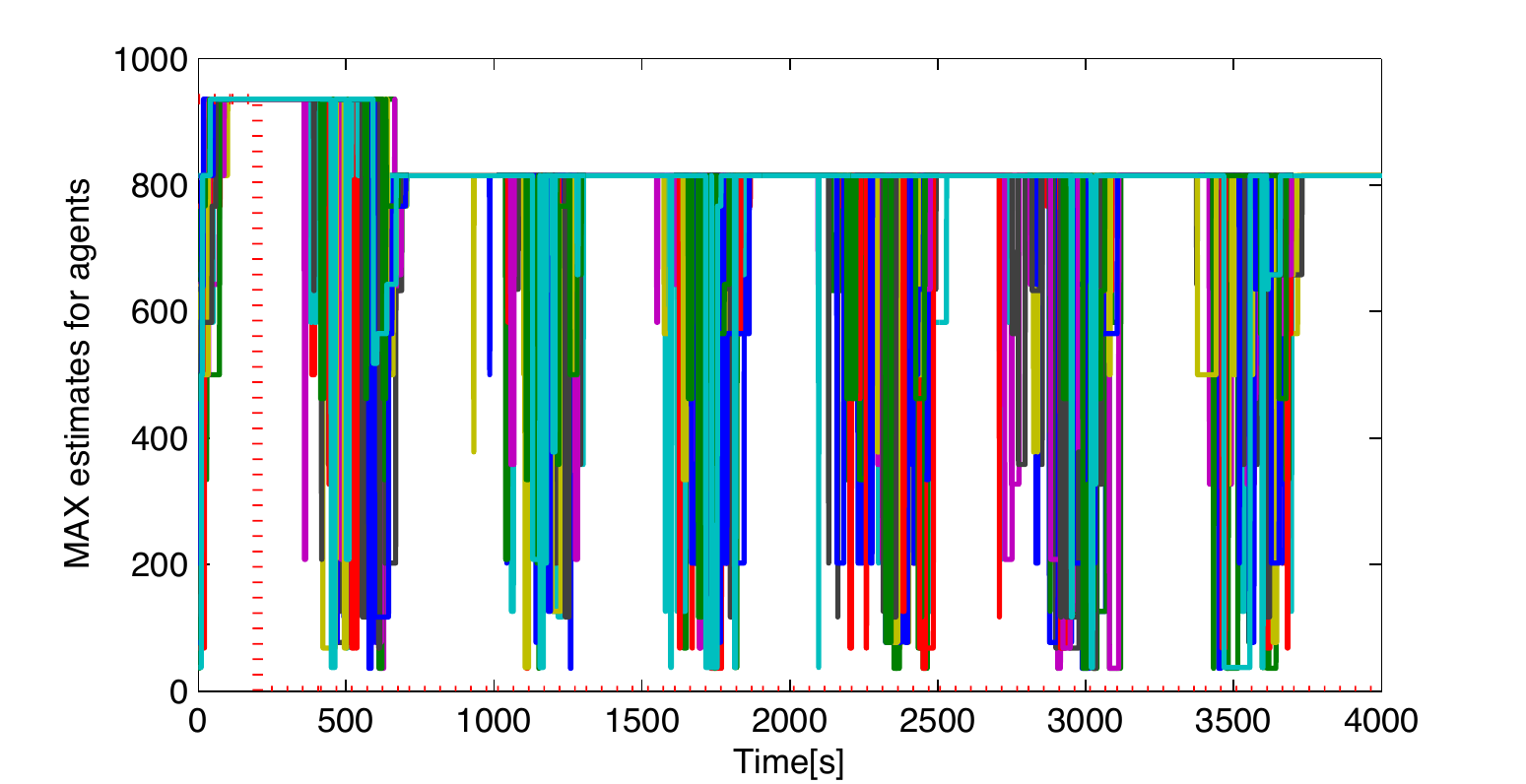}\\
(b)\\
\includegraphics[scale=.42]{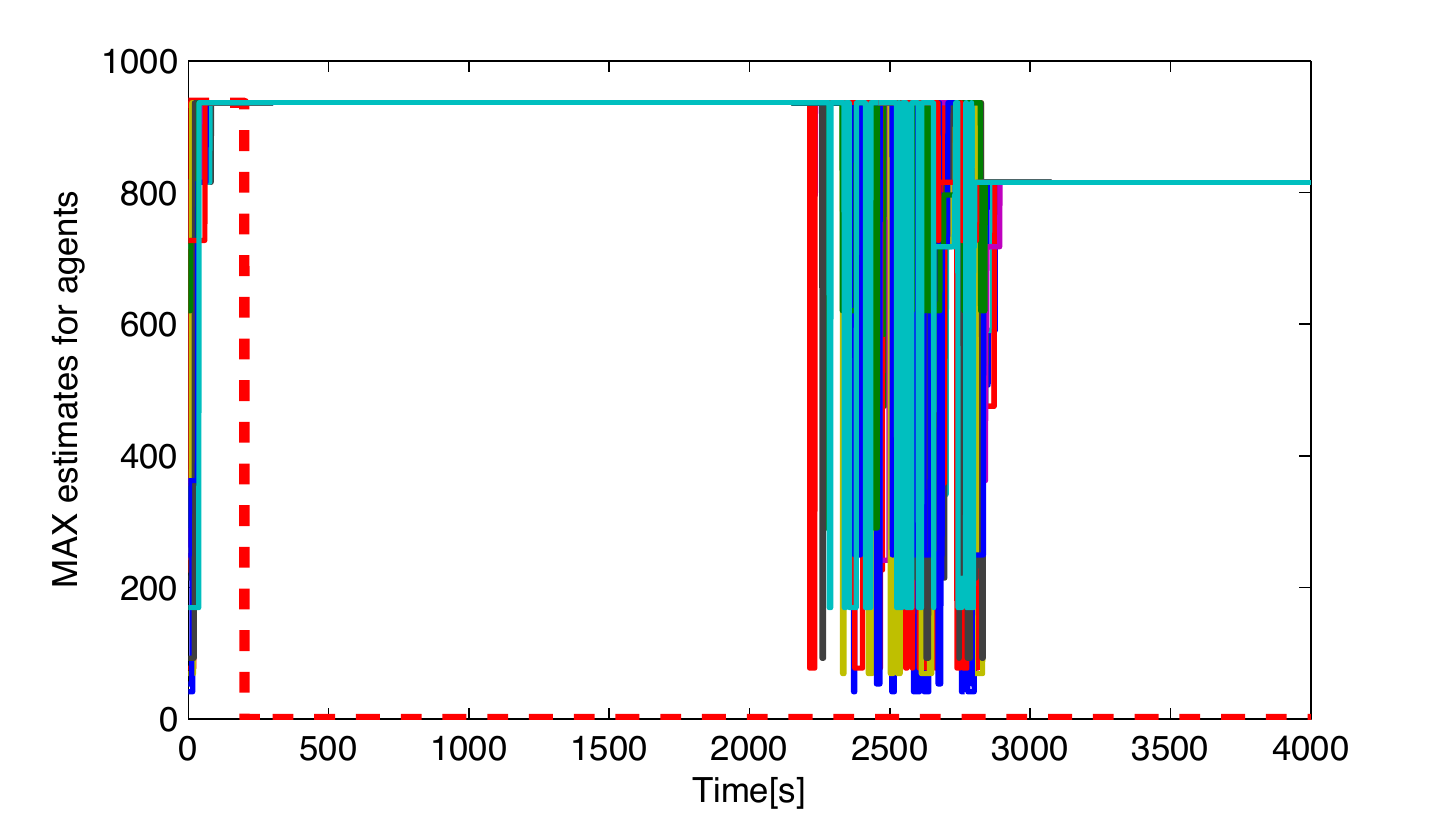}\\
\\
(c)
\end{tabular}
\caption{Evolution with time of the agent estimates $y_i(t)$ for the algorithm of Section \ref{sec: counters} where departures can be announced (a), and of Section \ref{sec: timeout} where departures are not announced, for tresholds $\T^*=40$ (b) and $\T^*=200$ (c). (The scale is different in (a)). Departure of the agent with highest value is represented by a dashed line. }
\label{fig:simulations}

\end{figure}

We have simulated the two algorithms, with two thresholds $\T^*$ for that of Section \ref{sec: timeout}, and the results are represented in Fig. \ref{fig:simulations}. We note that in the three cases,  all the agents first converge to the MAX value of $x_{9} = 936$ in a bit more than 100 time steps, before agent 9 leaves the network. After the departure of agent $9$ at $t=200$, we see that  the algorithm of Section \ref{sec: counters} that uses messages from departing agents reconverges to the new maximal value $x_{13} = 815$ in 137 time steps. The performance of the algorithm of Section \ref{sec: timeout} without messages from leaving agents are significantly worse. For a threshold $\T^*=40$, we see that it takes 506 time steps to reconverge to the new maximal value, but the system later suffers from several spurious resets. These are caused by agents reaching the threshold by chance. The probability of this occurring can be significantly reduced by taking a higher threshold, but this results in an even longer time to react to the departure of $9$, as seen in Fig. \ref{fig:simulations}(c) with $\T=200$.  2439 time-steps are indeed needed to re-obtain the correct value, mostly because it takes very long before the agents abandon their former estimate. This clearly illustrates the trade-off on the threshold value $\T^*$: a too small value will result in spurious resets as soon as some agents \quotes{have not heard} about the agent with the highest value for too long. But a too large threshold will result in a significant delay before agents decide that an agent has probably left the system.

We also performed comparisons between the two approaches on the convergence time to reach consensus after the agent with MAX has left the group for other numbers of nodes. The results are summarized in Table \ref{tble: comparison-number-nodes}. We take $\T^* = 1.1 \mathcal{N}(0)$ with the algorithm of Section \ref{sec: timeout}. We observe that when the number of agents increases, the algorithm of Section \ref{sec: counters} requires proportionally much fewer iterations to reach consensus, as expected and already observed in Fig. \ref{fig:simulations}. Moreover, it also achieves a stronger version of the property of eventual correctness than the algorithm of Section \ref{sec: timeout}, as it avoids spurious resets, as discussed above. It does however require the possibility of sending messages when leaving. 
\begin{table}[H]
\centering
\begin{tabular}{c|cc}
\toprule
\multirow{2}{*}{Number of nodes}  & \multicolumn{2}{c}{Iterations to reach MAX-consensus} \\
\cmidrule(r){2-3}
& Algorithm \ref{Intialization algorithm}-\ref{Modified random gossip} & Algorithm \ref{Intialization algorithm}, \ref{Joining algorithm_timeout}-\ref{Modified_random_gossip_timeout_sep_from_update} \\
\midrule
10 & 21 & 64 \\
20 & 129 & 162 \\
30 & 194 & 599 \\
50 & 246 & 1885\\
100 & 628 & 6580\\
\bottomrule
\end{tabular}
\vspace{0.3cm}\caption{Comparison between the two techniques with different number of nodes.}
\label{tble: comparison-number-nodes}
\end{table}

\section{Discussion and Conclusion}\label{sec: conclusion}

We have investigated the distributed MAX-consensus problem for \textit{open} multi-agent systems. Two algorithms have been proposed depending on whether the agents who leave the network can inform another existing agent about their departure or cannot. The eventual correctness has been proven for both.

Taking a step back, we see two main challenges in the design of algorithms for open multi-agent systems, as also briefly noted in \cite{Hendrickx-AAC16-Open}:

\emph{Robustness and dynamic information treatment:} The algorithms should be robust to departures and arrivals, in the sense that they should \MA{keep updating their estimates to discard outdated information.} Moreover, novel information held by arriving agents should be taken into account, and outdated information, for example, related to agents no longer in the system, should eventually be discarded.

\emph{Performance in open context:} The performance of classical multi-agents algorithms is often measured by the rate at which they converge to an exact solution or a desired situation (or the time to reach such a situation). This approach is no longer relevant in a context where agents' departures and arrivals keep \quotes{perturbing} the system, and possibly the algorithm goal (as is the case here). Rather, efficient algorithms would be those for which the estimated answer remains \quotes{close} to some \quotes{instantaneous exact solution}, according to a suitable metric.

The algorithms we have developed here do answer the first issue of robustness and information treatment for the problem of distributed maximum computation. 
The characterization and optimization of their performance in an open context, however, remains unanswered at present and could be the topic of further works.  We note that the behavior of a gossip averaging algorithm in an open multi-agent system was characterized in \cite{Hendrickx-AAC16-Open}, but this algorithm was not designed to compute a specific value, as is the case here.

In particular, we observe that both algorithms would suffer from occasional apparently unnecessary resets. This may happen after the departure of an agent that did not have the largest value in the algorithm of Section \ref{sec: counters}, or when an agent has been isolated for too long from that with the highest value in the algorithm of Section \ref{sec: timeout}. We do not know at this stage if these spurious resets can be entirely avoided, especially when leaving agents cannot send a final message. In this case, it is indeed impossible to know for sure whether the agent with the highest value has left or has just not communicated for a while. There are, however, several possibilities to mitigate the damage of these spurious resets and to play on the trade-off between the effect of these perturbation and the speed at which the system reacts. A simple solution could be for example to apply an additional filtering layer when the algorithm requires an important decrease of $y_i$. In this case, a new estimate $\tilde y_i$ would follow $y_i$ except that sharp decrease would be replaced by gradual ones.
We also observe that our second algorithm will either only work when the system size is not too large with respect to $\T^*$  (case of a fixed threshold) or eventually work for all size but gradually become slower and slower to react (case of a growing $\T^*$). Whether this can be avoided in a context when leaving agents do not warn others about their departure also remains an open interesting question.

\bibliographystyle{IEEEtran}


\end{document}